# Speckle-free laser imaging

Brandon Redding[1*], Michael A. Choma[2,3†], Hui Cao[1†]

[1]Department of Applied Physics, Yale University, New Haven, CT 06511, USA
[2]Departments of Diagnostic Radiology and of Pediatrics, Yale School of Medicine, New Haven, CT 06520, USA
[3]Department of Biomedical Engineering, Yale University, New Haven, CT 06511, USA

[†]These authors contributed equally to this work.
*E-mail: brandon.redding@yale.edu

Many imaging applications require increasingly bright illumination sources, motivating the replacement of conventional thermal light sources with light emitting diodes (LEDs), superluminescent diodes (SLDs) and lasers. Despite their brightness, lasers and SLDs are poorly suited for full-field imaging applications because their high spatial coherence leads to coherent artifacts known as speckle that corrupt image formation[1, 2]. We recently demonstrated that random lasers can be engineered to provide low spatial coherence[3]. Here, we exploit the low spatial coherence of specifically-designed random lasers to perform speckle-free full-field imaging in the setting of significant optical scattering. We quantitatively demonstrate that images generated with random laser illumination exhibit higher resolution than images generated with spatially coherent illumination. By providing intense laser illumination without the drawback of coherent artifacts, random lasers are well suited for a host of full-field imaging applications from full-field microscopy[4] to digital light projector systems[5].

Lasers are indispensable light sources in modern imaging systems. Intense laser sources enable imaging through scattering or absorptive media, or imaging dynamic behavior on short time scales. One of the signature properties of conventional lasers is high spatial coherence, a property resulting from resonant cavities with a limited number of spatial modes that produce well-defined wavefronts. A high-degree of spatial coherence has well-known advantages and disadvantages. On the one hand, high spatial coherence allows for the highly directional emission of traditional lasers. On the other hand, spatial coherence leads to coherent imaging artifacts such as speckle. Speckle is generated when an imaging system (e.g. optical aberrations, surface defects) or sample (e.g. scattering) imparts a range of random path length differences in mutually coherent photons that subsequently interfere at a detector[6]. This interference causes artificial modulations in the measured intensity, which degrades the image quality. Researchers have sought methods to mitigate the effects of laser speckle[7] by "scrambling" the laser wavefront with a moving phase plate and averaging the time-varying speckle pattern; but this requires long integration time as the visibility of the speckle only decays as $M^{-1/2}$, where $M$ is the number of independent speckle patterns[8].

Random lasers are an unconventional laser in that they are made from disordered materials that trap light via multiple scattering[9, 10]. The spatial modes are inhomogeneous and highly irregular. With external pumping, a large number of modes can lase simultaneously with uncorrelated phases. Their distinct and richly structured wavefronts combine to produce spatially incoherent emission. Our recent studies show that the spatial coherence of random laser emission from a laser dye solution (5 mM rhodamine 640 dissolved in diethylene glycol) interspersed with scattering particles (240 nm diameter polystyrene spheres) can be controlled by adjusting the scattering strength and the pump geometry. Based on this finding, we are able to engineer the random laser to achieve low spatial coherence. For example, when the suspension has a scattering mean free path of 100 μm and the pump light is focused to a spot of diameter 275 μm, the mutual coherence of the laser emission at a spatial distance of 125 μm is less than 0.1.

To compare the brightness of a random laser to the existing light sources used for imaging applications, we evaluate the photon degeneracy, $δ$, a common metric used to evaluate the relevant intensity of a light source. Photon degeneracy is defined as the number of photons per coherence volume, where the coherence volume is the product of the coherence length (temporal coherence) and coherence area (spatial coherence)[11]. Thermal sources and LEDs have very low photon degeneracy. For a thermal source, $\delta$ depends on the temperature and is $\sim 10^{-4}$ at 3000 K[11]. A high efficiency LED has $\delta$ on the order of $10^{-3}$[12]. SLDs and conventional lasers, both exhibiting high spatial coherence, have photon degeneracy much larger than 1. For a typical SLD, $\delta$ is estimated to be $\sim 10^3$ [13]. Conventional lasers not only produce intense radiation, but also have a large coherence volume, leading to extremely high photon degeneracy: a typical, single-mode HeNe laser emitting 1 mW has $\delta \sim 10^9$ [11]. We estimated the photon degeneracy for the random laser used in this work from the coherence volume of emission, $\Delta V$, and the number of photons per unit volume, $\rho$, as $\delta = \rho \Delta V$. Due to the low spatial and temporal coherence of the random laser, $\Delta V$ is much smaller than that of a conventional laser. In our experiment, the dye molecules were optically excited with a frequency doubled Nd:YAG laser operating at 532 nm with 30 ps pulses and a repetition rate of 10 Hz. During the $\lesssim$100 ps emission pulse, the peak random laser power was ~530 W, yielding an average power of 530 nW at our 10 Hz repetition rate. Due to the fast radiative decay rate of the rhodamine 640 dye molecules, the pump rate could be increased to ~MHz without changing the random laser performance and we would expect the average power to scale accordingly. The random laser degeneracy with the 10 Hz pump laser is $\sim 1.7 \times 10^{-2}$. If the repetition rate is increased to MHz, we would expect $\delta$ to increase to $\sim 10^3$. From this analysis, it is clear that a random laser can provide many orders of magnitude improvement in photon degeneracy compared with a thermal light source or an LED. As illustrated in Fig. 1, this combination of high photon degeneracy and low spatial coherence has not been realized in other light sources and makes random lasers uniquely suited for imaging applications.

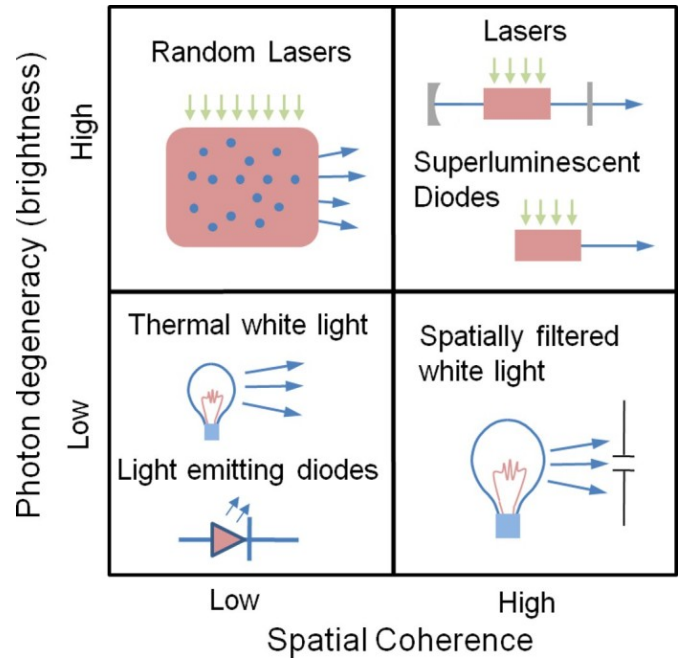

**Fig. 1. Light sources for imaging.** Light sources are compared in terms of the two parameters most relevant to full-field, non-scanning imaging: the photon degeneracy and the spatial coherence. Random lasers represent a new class of light source with high photon degeneracy and low spatial coherence—the ideal combination for full-field imaging.

To demonstrate that the low spatial coherence of a random laser does in fact enable speckle-free imaging, we compared images generated with spatially incoherent random laser illumination to those generated with spatially coherent amplified spontaneous emission (ASE) illumination. While a dye solution containing scattering particles (of concentration $6.1 \times 10^{12}$cm$^{-3}$) was optically excited to produce random laser emission, the ASE (with similar intensity and spectral width) was obtained from the same kind of dye solution without scatterers. A Young's double slit experiment was conducted to confirm that the spatial coherence of the ASE was much higher than that of the random laser emission[3].

First, we demonstrate speckle suppression under random laser illumination by comparing the speckle patterns generated by illuminating a scattering medium with the two sources. The scattering medium consisted of a 3 μm thick film of $TiO_2$ particles spun onto a glass coverslip. The particles were ~20 nm in diameter and the transport mean free path was ~600 nm. Images of the two light sources transmitted through the scattering medium are shown in Fig. 2(a,b). While speckle is clearly visible with the spatially coherent ASE illumination, it is greatly suppressed by the random laser illumination. As a quantitative comparison, we extracted the probability, $P$, of finding a pixel with a given intensity, $I$, normalized by the average intensity, $I_0$, of all the pixels. This probability density function is plotted in Fig. 2(c). The relatively narrow intensity distribution under the random laser illumination is contrasted with the broad distribution in the ASE illumination.

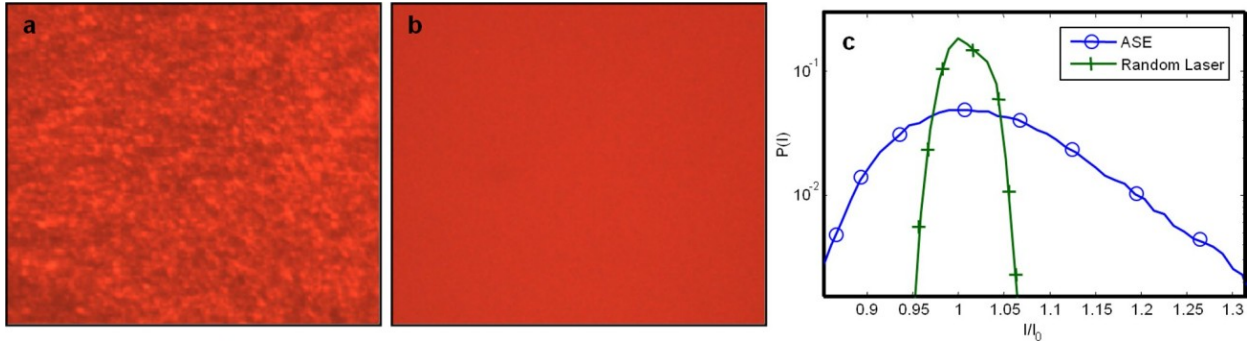

**Fig. 2. Speckle suppression.** (a,b) Speckle images generated by a thin scattering film illuminated with spatially coherent ASE (a) and by spatially incoherent random laser emission (b). (c) Intensity fluctuations in the images (a,b) measured by the probability density function of light intensity $I$ at each pixel of the camera, normalized by the average intensity $I_0$ of all pixels.

We then demonstrate improved imaging resolution under random laser illumination without a scattering medium. In the imaging setup, the random laser or ASE emission illuminated a 1951 US Air Force (USAF) resolution test chart, which was imaged in transmission mode onto a camera (Moticam 2300) by a spherical lens (numerical aperture =0.42). We collected images using both critical and Köhler illumination and obtained similar results. The images with the ASE and random laser illuminations are shown Fig. 3(a,b). Note that the parameters of the camera (i.e. integration time, gain level) were identical in both cases. Due to spherical aberrations of the imaging lens, the features on the test chart appear slightly blurred. Under spatially coherent ASE illumination, interference occurs in the regions where these blurred features overlap, further degrading the image. To quantitatively compare the resolution of these images, we extracted cross sections of the group 6 features [vertical lines in Fig. 3(a,b)]. As shown in Fig. 3(c), the visibility of these features, defined as $(I_{max}-I_{min})/(I_{max}+I_{min})$, decreases for smaller features, as expected. However, the visibility reduction is clearly more dramatic under the ASE illumination than the random laser illumination. In Fig. 3(d), we plot the visibility as a function of spatial frequency, and confirm that the imaging resolution is improved by using random laser illumination.

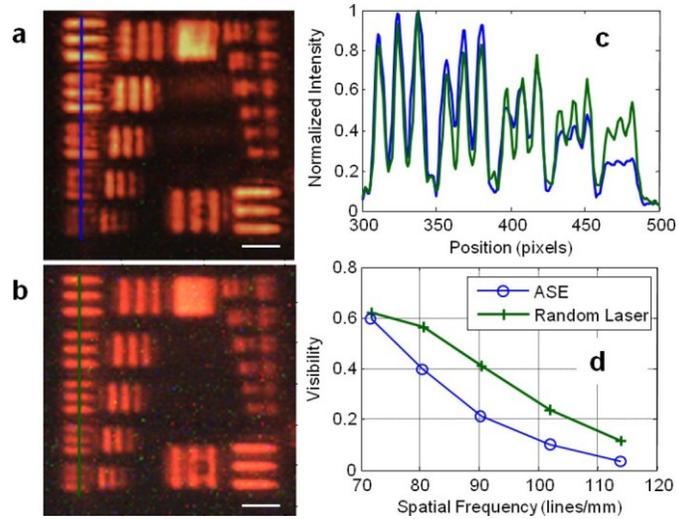

**Fig. 3. Image comparison without a scattering medium.** (a,b) Images of an Air Force resolution test chart with spatially coherent ASE illumination(a) and spatially incoherent random laser illumination (b). The scale bars are 30 μm. (c) Cross section of the features on the left side of the images indicated by the green and blue lines in (a,b). (d) Visibility of the features in (c) as a function of their spatial frequencies.

The benefit of imaging with a spatially incoherent source is even more pronounced in the presence of a scattering medium which introduces additional opportunities for coherent artifacts. We imaged the USAF resolution test chart through the same $TiO_2$ scattering film used to demonstrate speckle suppression above. The images of the chart illuminated with the random laser and the ASE are shown in Fig. 4(a,b). In comparison with the images in Fig. 3(a,b), we observe that the scattering film effectively increased the background signal because photons were scattered to what would otherwise be dark regions of the image. Under ASE illumination, interference among these scattered photons resulted in artificial features which corrupt the image nearly beyond recognition. In contrast, random laser illumination generates a clear image, demonstrating the advantage of imaging with a spatially incoherent source. For a quantitative comparison of the two images, we calculated the contrast to noise ratio (CNR) for each set of features with different spatial frequencies. The CNR is defined as $(<I_s> - <I_b>)/\sigma$, where $<I_s>$ and $<I_b>$ are the average intensities in the pattern (bright area) and the background (dark area), and $\sigma$ is the standard deviation of the intensity in space. As shown in Fig. 4(c), the image collected with random laser illumination exhibited higher CNR at all spatial frequencies.

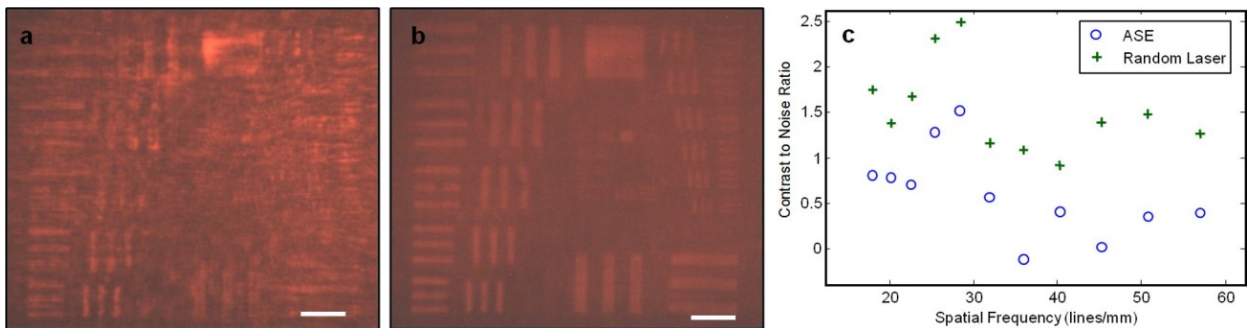

**Fig. 4. Image comparison with a scattering medium.** (a,b) Images of a USAF resolution test chart through a scattering film with spatially coherent ASE illumination (a) and with spatial incoherent random laser illumination (b). The scale bars are 50 μm. (c) Contrast to noise ratio (CNR) extracted from the images (a,b) showing the improved image quality with random laser illumination.

In conclusion, we demonstrated that random lasers, with low spatial coherence and high photon degeneracy, represent a new class of illumination source that is ideal for a wide range of full-field imaging applications. In addition to their low spatial coherence, random lasers typically exhibit low temporal coherence owing to their broad emission spectra. The temporal coherence length of the random laser considered in this work, for instance, can be estimated from the 10 nm emission bandwidth to be ~ 17 μm[14]. This short temporal coherence could allow random lasers to be used in coherent imaging applications such as optical coherence tomography [15], which are also known to suffer from spatial coherence induced artifacts[16, 17].